# Accepted Manuscript

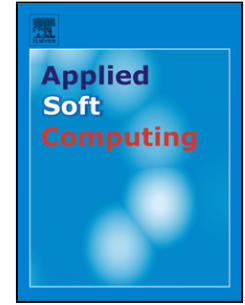

Title: Advanced Monitoring of Rail Breakage in Double-Track Railway Lines by means of PCA Techniques

Authors: F. Espinosa, J.J. García, A. Hernández, M. Mazo, J. Ureña, J.A. Jiménez, I. Fernández, C. Pérez, J.C. García

# Advanced Monitoring of Rail Breakage in Double-Track Railway Lines by means of PCA Techniques

F. Espinosa, J. J. García, A. Hernández, M. Mazo, J. Ureña, J. A. Jiménez, I. Fernández, C. Pérez, J.C. García.

Electronics Department, University of Alcala

E-28801 Alcalá de Henares (Madrid), Spain

Graphical abstract

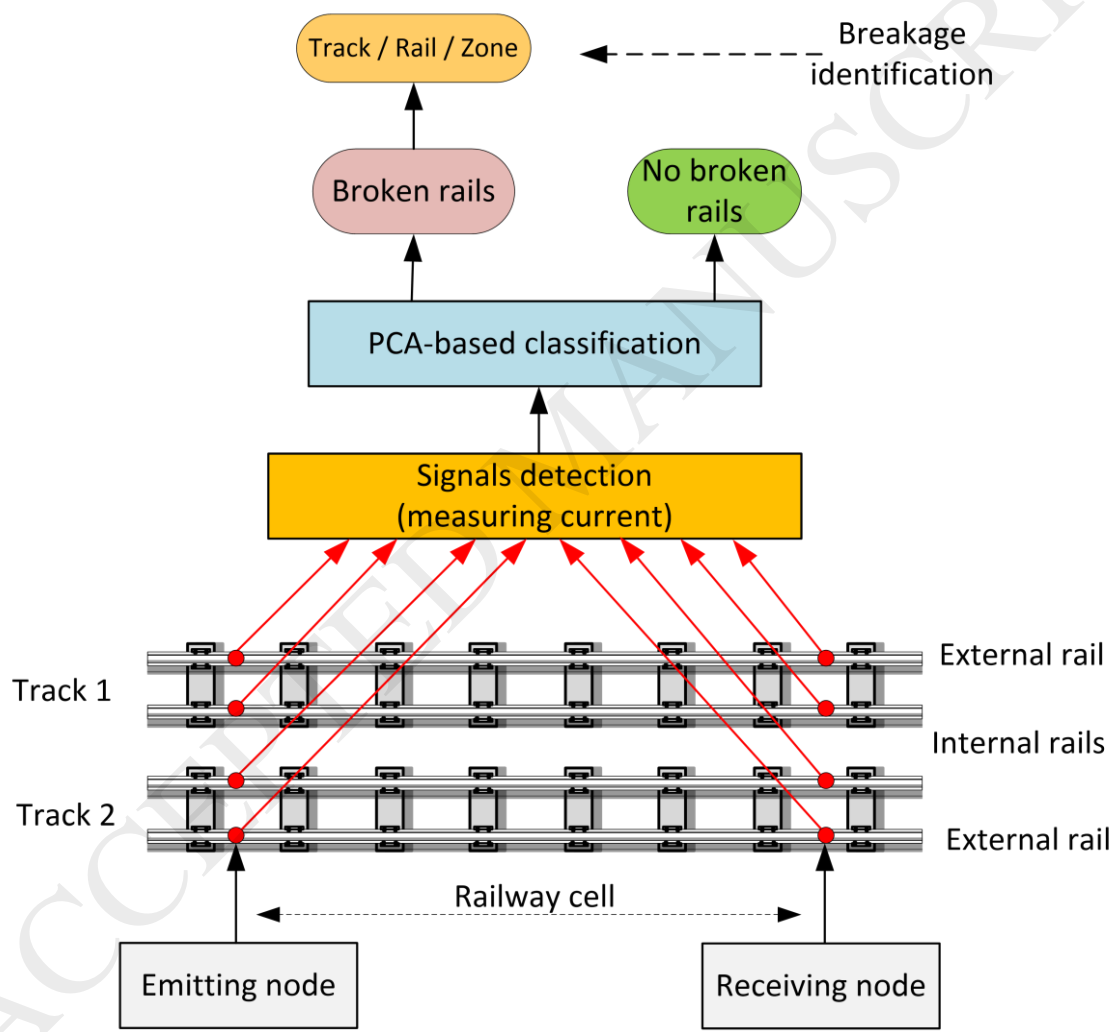

**Highlights**

- New strategy or rail breakage detection for a double-track railway line.
- Outputs: breakage railway identification, broken track detection, broken zone detection.
- Experimental results show a 100% success rate identifying the breakages.
- The results show that the proposed system helps to have a safety control of the railway traffic.

**Abstract**

This work describes a classifier designed to identify rail breakages in double-track railway lines, completing the electronic equipment carried out by authors. The main objective of this proposal is to guarantee the integrity of tracks before the railway traffic starts working. In addition, it facilitates maintenance tasks providing information about possible breakages. The detection of breakages is based on the analysis of eight currents provided by the electronic equipment, one per rail, at the ends of the section (emitting and receiving nodes). The imbalance that occurs among the value of these currents implies that there is at least a breakage in the track section under analysis. This analysis is conducted according to three phases. The first one identifies whether there is a breakage, and, in that case, the damaged track is identified. The second phase provides information about which rail is broken (internal, external or both of them) in the previously identified track. Finally, if there is only one breakage, the third phase estimates its most likely zone along the track section. This situation is considered as a classification problem, and solved by means of the Principal Component Analysis technique. This means that a significant number of measurements is required for every breakage pattern (types of breakages) to be considered. Due to the difficulty of having real data, the proposal has been validated using an 8km-long double-track hardware simulator specially designed by the authors, with specific localizations for breakages.

## 1. Introduction

The detection of broken rails in railway environments has been widely studied during the last decade. The early broken rail detection has become a critical issue especially in high-speed lines, not only for safety reasons but also for the efficient maintenance [1]-[4]. For that purpose, several approaches have been proposed, depending on the sensory technology applied, as well as different techniques [5]. An extended overview about technologies regarding broken rail detection can be found in [6][7]. Particularly we would like to highlight the systems based on non-destructive testing (NDT) [8]-[12] and track circuit [13]-[19].

NDT approaches are often based on ultrasounds to explore railways [11][12][20]. They allow to detect not only breakages, but also cracks or superficial damage. Nevertheless, they are frequently characterized by a complex infrastructure to be installed along the tracks, as well as a limited maximum distance (around 2 km) to be scanned by a single setup. Optic fiber has been also installed along the railways [21], in order to detect broken rails [22]. Although this approach provides flexible and easily installable solutions, it can be rapidly degraded due to the high strain supported by rails [23]. The most extended track circuits as breakage detector are the ones based on AC audio-frequency [19]. They achieve distances around 2km, what means a considerable number of electronics systems along the track to properly monitor long railway lines, thus increasing installation and maintenance costs.

In addition to the technologies, some commercial proposals are also described in [7], concluding that there is a paucity of literature with regard to reliable solutions for high speed broken rail detection in real-time, being the ultrasonic-based method the most common inspection technique for broken rail. One of the few automated commercial solutions available on the market is the Ultrasonic Broken Rail Detector (UBRD) from RailSonic [24]. However, the section coverage is limited up to 2.5km, requiring a monitoring system every 1.75km.

Other issue arises when sensory data are analyzed to determine whether there is or not a breakage along the track, and where it is located. For that purpose, not only sensor data can be considered, but also any previous information about the system that could increase reliability. This is the reason why some works propose high-level algorithms in order to make fusion and decisions about the rail status over time. In [25] a Bayesian network is designed for the diagnosis of a railway, especially for the classification of defects. Some authors propose the use of different classifiers for the fault diagnosis in railway track circuits, such as neuro-fuzzy systems [26], Dempster-Shafer theory [27], or support vector machine (SVM) [28]. Neural networks and SVMs have been also analyzed and compared in [29], for monitoring wooden railway sleepers. Other authors propose the use of Principal Component Analysis (PCA) technique for solving the classification problem, which has been applied in numerous works to extract and classify information coming from different sensory systems [30]. Examples of use of PCA can be found as classification process with ultrasonic systems [31], as voice and character recognition [32], as machine defect classification [33], or face representation and recognition [34]. PCA is also a useful technique in the fields of fault detection and diagnosis [35][36][37].

In this context, the authors have proposed in a previous work [38] an electronic system for the real-time simultaneous detection of breakages in single- or double-track railway lines. The approach is based on the electrical discontinuity of one or more rails in a double track railway, always assuming that rails are electrically insulated from the railway infrastructure at the working frequency (800Hz). This electronic system provides simultaneously the measurement of 8 currents in different tracks and localizations (the distance between some of the measured currents can be more than 8 km). If the analysis of these currents is considered as a classification problem, not only does it help to identify the broken rail, but also to know which the corresponding track is and to estimate the zone in which the breakage is located. For solving the classification problem, we propose a high-level algorithm based on Principal Component Analysis to complete the previous developed electronic system.

The remainder of the manuscript is organized as follows: Section 2 describes the background and an introduction to the developed double track hardware simulator; the PCA-based broken rail detector is detailed in Section 3; Section 4 presents experimental results for evaluating the proposed solution; Section 5 is dedicated to the discussion of the results and provides some directions for future work; and, finally, Section 6 concludes the paper.

## 2. Background

In a first approach, we could evaluate each rail independently, but it would require a new electric cable parallel to the railway as returning circuit for the current, what would be part of the solution and source of new problems. This is why we bet on a system that does not require more electrical conductors than the own rails and we take advantage of the magnetic coupling among rails, that generate different current imbalance depending on the breakage location. The two main contributions of the authors' proposal, compared to previous ones, are: a) the increase of the distances between nodes in double-track lines, thus implying a shorter time for analyzing the whole railway line; and b) the advantage of approximately detecting the breakage location.

As we propose the use of PCA for classifying the different breakages, next subsections introduce the set of currents that will be used as feature vector, and the hardware platform for testing the classifier.

**2.1 Feature vector**

The active broken rail monitoring system has been designed for a high-speed railway with two tracks. The train is powered through the catenary and the pantograph, and the global electric power is distributed in electrically isolated sections, each one associated to an electrical substation. The voltage is kept constant along the section by means of auto-transformers. The section between two transformation centers is typically from 10km to 15km long. Fig. 1 shows the configuration proposed to measure the electrical continuity of rails in a section with two tracks. It includes emitting nodes at the transformation centers, whereas a receiving node is placed in the middle. The electronic system and measurement methodology designed by the authors is described in detail in [38]. The emitted currents are modulated with Kasami codes and the detection is carried out at the receiving nodes by using correlation techniques. Considering the distance between transformation centers, where the electronic system would be installed, this encoding scheme provides two important advantages: a significant process gain and a high immunity to noise. Furthermore, due to the suitable cross-correlation properties of these codes, simultaneous emissions can be carried out from every emitting node, being possible to distinguish the origin of each transmission at the receiving node.

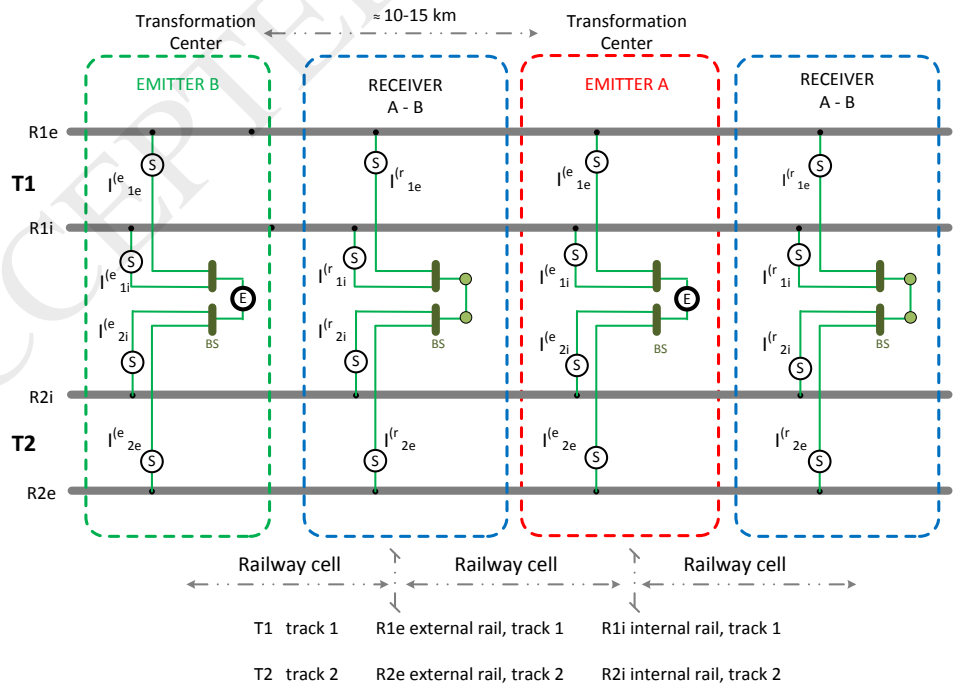

T1 track 1     R1e external rail, track 1     R1i internal rail, track 1
T2 track 2     R2e external rail, track 2     R2i internal rail, track 2

Fig. 1. Configuration proposed to detect breakage (electrical discontinuity) in a double-track railway line. Locations of the emitting and receiving nodes between two transformation centers.

Note that each emitting node encodes its emission with a different code (A or B) whereas the receiving node has to detect both of them (see Fig. 1); in this way the breakage testing time is significantly reduced. Both kinds of nodes measure the current associated to each rail. Then, considering an emitter-receiver pair, eight currents $I_{tr}^{(n)}$ are available according to (1) which complete the 8-dimensional *feature vector* (2) to be used in the classifier. The emitting node is the same for two consecutive cells, so the currents $I_{tr}^{(n)}$ are simultaneously measured in both of them.

$$I_{tr}^{(n)} \quad ; \quad \begin{aligned} n &= \begin{cases} e & \text{measured at emitter} \\ r & \text{measured at receiver} \end{cases} \\ t &= \begin{cases} 1 & \text{measured at track no. 1} \\ 2 & \text{measured at track no. 2} \end{cases} \\ r &= \begin{cases} e & \text{measured at external rail in the corresponding track} \\ i & \text{measured at internal rail in the corresponding track} \end{cases} \end{aligned} \quad (1)$$

$$\dot{x} = \begin{bmatrix} I_{1e}^{(e} & I_{1i}^{(e} & I_{2i}^{(e} & I_{2e}^{(e} & I_{1e}^{(r} & I_{1i}^{(r} & I_{2i}^{(r} & I_{2e}^{(r} \end{bmatrix} \quad (2)$$

The electronic system implemented to measure and process these currents (2) provides two different ways of generating the signal injection: either independently for each track *t*, or jointly for both tracks [38]. In the first case, the modulated signal is applied to each track by using a time division multiplexing (see Fig. 2a). This means that four independent currents are obtained for each track. However, in the joint injection, the modulated signal is applied between both tracks by means of the suitable connection, as is shown in Fig. 2b. In this second case, eight currents are simultaneously measured, existing a correlation among them. These two injection modes play a significant role in the detection and location of rail breakages as it is discussed in the following section.

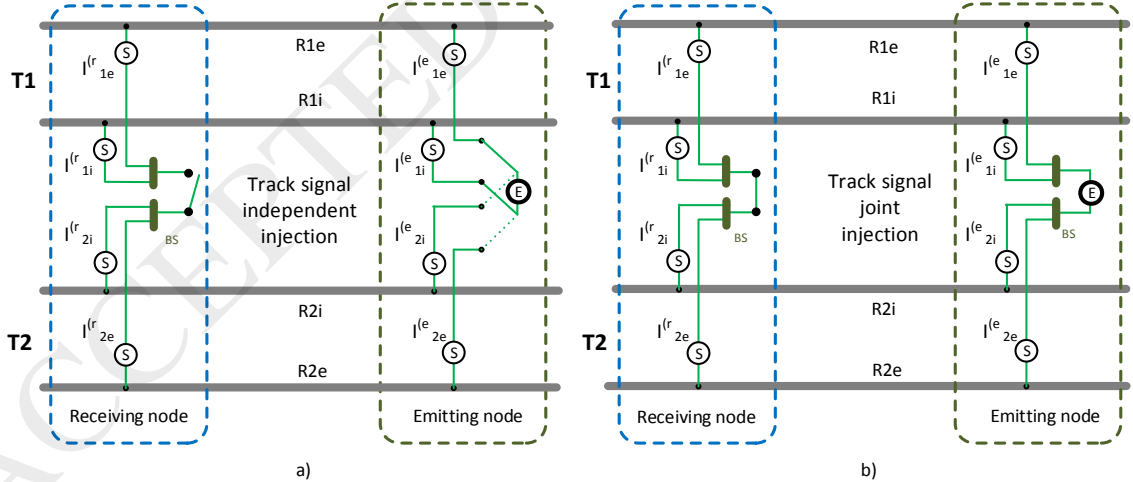

Fig. 2. Configuration at the emitting and receiving nodes according to the signal injection mode: a) independent injection; b) joint injection.

### 2.2. Hardware simulator

Although broken rails are the leading cause of major derailments accidents [39][40], it is very difficult to get enough evidences to validate a proposal as the one described in this paper. For that reason the authors have developed a double track hardware simulator, based on the impedance of the high speed tracks [41]. The objective of this simulator is to generate track breakages [38] in different positions of the track section. It includes four impedance modules equivalent to 2km-long track sections in dry conditions (water content is 0.1%) and at a working frequency of 800Hz. Every module is composed of a set of discrete RLC. Their serial connection provides the electrical characteristic of a track section between the emitting and receiving nodes, with a length of 8km. This system allows to simulate a breakage in any rail at three different positions: near the emitting node (a quarter of the track length, 2km), near the receiving one (three quarters of the track length, 6km), or in an intermediate area (half the track length, 4km). Fig. 3 shows the block diagram of the developed system, the possible location of the breakages (breakage switches) along the track section under supervision and the nodes where the currents $I_{tr}^{(n)}$ belonging to the feature vector are measured (the emitting node on the left and the receiving one on the right). Besides, the figure includes the labels that will be used to identify every considered breakage corresponding to the third phase of the detection process described in Section 3.

The hardware simulator includes the electronic equipment for both nodes, providing the two types of signal injection: independently for each track and jointly for both tracks.

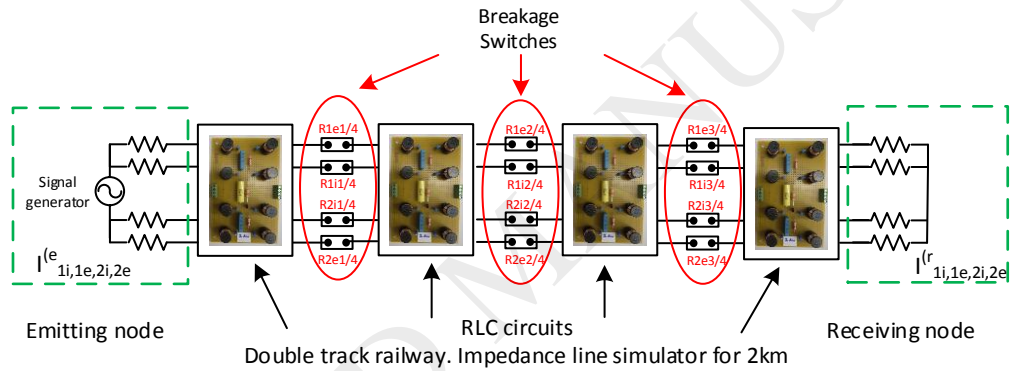

Fig. 3. Block diagram of the hardware simulator for a double-track section of 8km in length. Location of the breakages in the simulator.

## 3. Railway breakage detection process

The global idea of the detection system is shown in Fig. 4. By means of eight measurements (currents $I_{tr}^{(n)}$) we can supervise railway sections ended by an emitting node and a receiving one. The basic objective is to determine whether the railway cell between the emitting and receiving nodes is free of breakage or not. Nevertheless, in case of one breakage, our main challenge is to provide information about its location: close to the emitting node, close to the receiving one or in the intermediate area. To the best of our knowledge, this extra information has not been previously generated by any current broken rail detector. Moreover, considering the long distance between the emitting and receiving nodes, this approach can be very useful for maintenance operations.

Since there are different situations of breakage, each one is characterized by a class $\alpha_{t,r}$, where *t* represents the number of the track {1, 2} and *r* means if the breakage is on the external rail (e), on the internal one(i), or on both (ie). According to the measurements $I_{tr}^{(n)}$, the classifier reports the most likely breakage class $\alpha_{t,r}$.

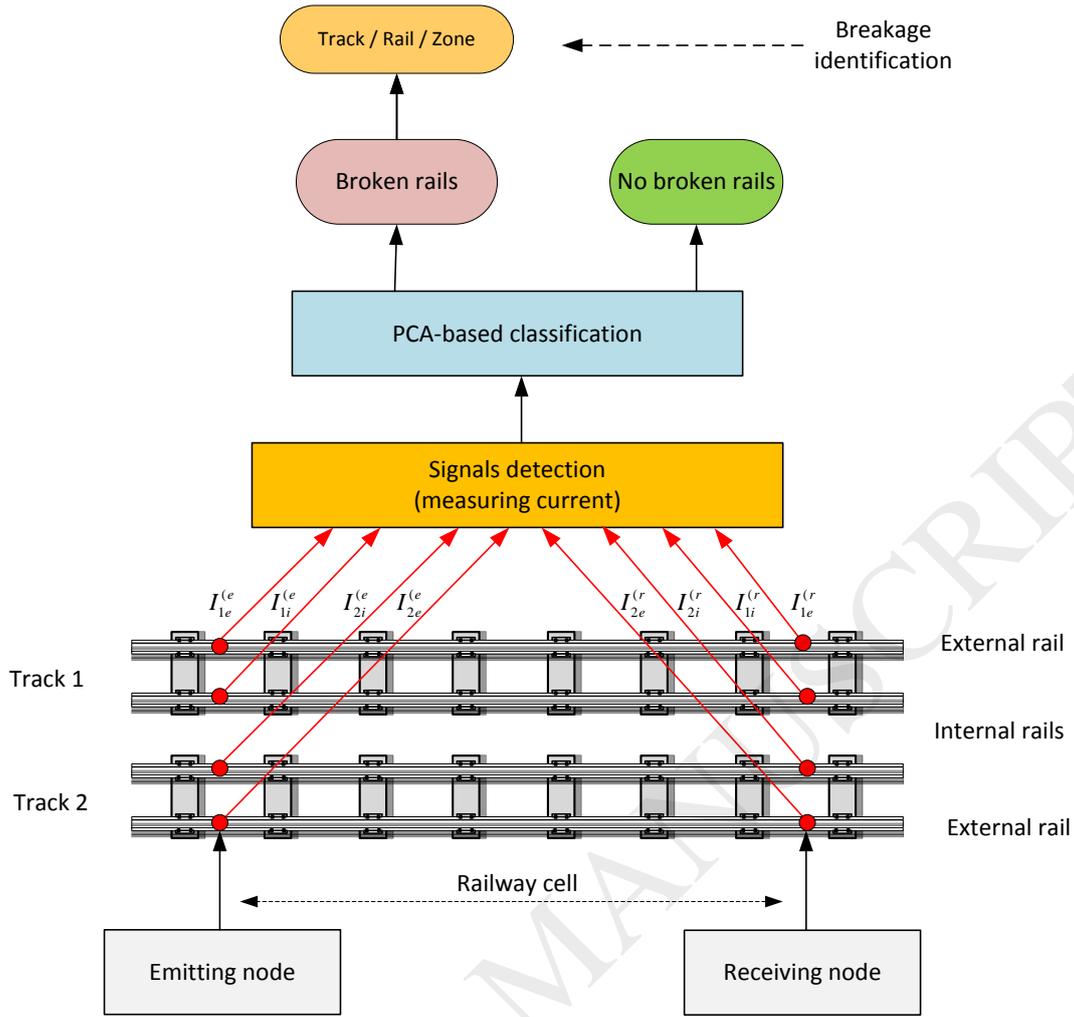

Fig. 4. Block diagram of the detection system.

The detection process is divided into the following phases:

- **Phase 1**. This first phase evaluates if there is a breakage in the railway cell under test. For doing that, the signal injection is carried out independently for each track *t* (see Fig. 2a). In this case four currents per track are measured (two in the emitting node and two in the receiving one). This information can be represented with two 4-dimensional feature vectors: $\dot{x}_1$ and $\dot{x}_2$ (3). As a result of this phase, the state of track *t* can be concluded: *correct* ($\alpha_{t,OK}$) or *broken* ($\alpha_{t,BR}$). Considering both tracks, four classes can be determined as Table 1 shows.

$$\dot{x}_1 = \begin{bmatrix} I_{1e}^{(e} & I_{1i}^{(e} & I_{1i}^{(r} & I_{1e}^{(r} \end{bmatrix}$$
$$\dot{x}_2 = \begin{bmatrix} I_{2e}^{(e} & I_{2i}^{(e} & I_{2i}^{(r} & I_{2e}^{(r} \end{bmatrix} \tag{3}$$

If the conclusion of the first phase is that there is only one broken track *t*, the detector continues with the next phase, otherwise the detection process finishes.

| Track breakage | Class | Broken track(s) |
|---|---|---|
| Track no. 1 {$t=1$} or Track no. 2 {$t=2$} (4 classes, 2 per track) | $\alpha_{t,OK}$ | Track $t$ is not broken |
| | $\alpha_{t,BR}$ | Track $t$ is broken |

Table 1. Classes related to the first phase of the detection process.

- **Phase 2**. This phase detects which the broken rail $r$ is: either the external, the internal or both. For this phase and the following one, the signal injection is carried out jointly for both tracks (see Fig. 2b), so an 8-dimensional feature vector $\dot{x}$ is obtained (2) for the cell under test. Table 2 shows the six classes $\alpha_{t,r}$ related to the second phase of the detection system.

| Track breakage | Class | Broken rail(s) |
|---|---|---|
| Track no. 1 {$t=1$} or Track no. 2 {$t=2$} (6 classes, 3 per track) | $\alpha_{t,i}$ | Track $t$, internal rail ($r=i$) |
| | $\alpha_{t,e}$ | Track $t$, external rail ($r=e$) |
| | $\alpha_{t,ie}$ | Track $t$, internal and external rails ($r=ie$) |

Table 2. Classes related to the second phase of the detection process.

If it is detected that both rails of the same track are broken the process finishes here (class $\alpha_{1,ie}$ or $\alpha_{2,ie}$). Otherwise the next phase is run.

- **Phase 3**. Assuming that the result from the previous phases is only one breakage (one track $t$, one rail $r$ and no more than one breakage per rail), the objective at this point is to estimate the position of this breakage along the track. Although three possible locations $z$ have been considered (close to the emitter $z=ne$, close to the receiver $z=nr$, or in the intermediate area $z=in$), the study could be extended to more positions without loss of generality. Considering both tracks, Table 3 shows the new twelve classes $\alpha_{t,r,z}$.

| Track breakage | Class | $\alpha_{t,r,z}$ Estimated position of the breakage |
|---|---|---|
| Track no. 1 {$t=1$} or Track no. 2 {$t=2$} (12 classes) | $\alpha_{t,r,ne}$ | Track $t$, rail $r$ {$i$ or $e$}, near the emitting node ($z=ne$) |
| | $\alpha_{t,r,nr}$ | Track $t$, rail $r$ {$i$ or $e$}, near the receiving node ($z=nr$) |
| | $\alpha_{t,r,in}$ | Track $t$, rail $r$ {$i$ or $e$}, in the intermediate area ($z=in$) |

Table 3. Classes $\alpha_{t,r,z}$ related to the third phase of the detection process.

Fig. 5 depicts the flow chart of the described detection process. According to it, if the third phase is reached it means that there are twelve different classes $\alpha_{t,r,z}$ to describe the situation of the railway cell as Table 3 shows. In this case the classes are sorted according to the position of the breakage.

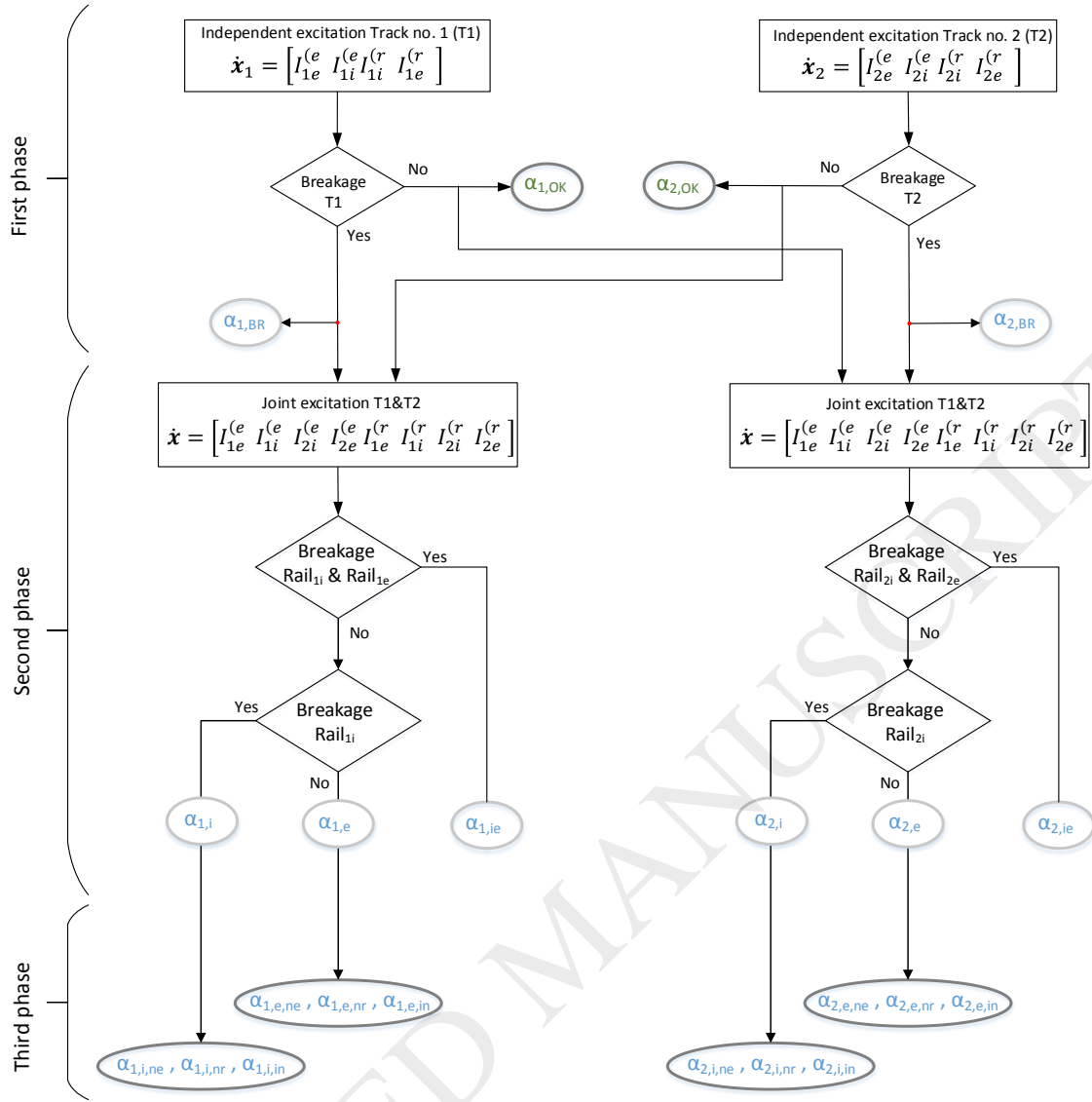

Fig. 5. Detection process: detection phases and classes.

## 4. PCA-based classifier

Once the measurements are available, the next step is to process them and to conclude which of the described classes is the most likely according to the value of the feature vector (distribution of the currents $I_{tr}^{(n)}$) following the described detection phases. For the first phase, the measurements of the four correlated currents per rail are used as input data (without correlation between both sets of currents since the independent injection is applied). If the obtained result entails to continue with the following detection phases, the input data is a set of 8 correlated currents (due to the joint injection) as is indicated in (2).

The existing correlation among the obtained currents at the second and third phases justifies the use of multivariate analysis in order to reduce the number of variables and to create the minimum set of uncorrelated ones. One approach to reach this objective is the use of PCA (Principal Component Analysis) [30].

Working with PCA two statistics are commonly used: Q statistic (squared prediction error, SPE) and Hotelling's T$^2$ statistic [35][37], for sensor fault detection as well as for classification applications. In our work we use the reconstruction error based on the Mahalanobis distance, which is similar to the Q statistic. With regard to the T$^2$ statistic, it measures the statistical variation due to systematic errors within the model of PCA. Since in this application the systematic error is negligible, this statistic is very low as we show in Section 6.

### 4.1. PCA technique overview

This technique transforms the input information from the original space to the PCA transformed space, with only the uncorrelated information. This transformation is carried out through the so-called transformation matrix $U$. If there are $j$ classes, one transformation matrix $U_j$ is calculated for each class $\alpha_j$, $j=\{1, …, c\}$. For the sake of clarity each class is defined in this Section as $\alpha_j$. The relationship between both notations is shown in Section 5. Without loss of generality, it is assumed the use of 8 currents, being the first phase a particular case (only four currents).

In general terms, PCA is divided into two stages. The first one, also called the training stage, is carried out off-line, when varying operational conditions have been taken into account, with the track section simulating one of the classes $j$. In this situation a data set $\dot{x}_j$ is captured, and the zero mean data set $x_j$ is used to obtain the transformation matrix $U_j$ between the original space and the transformed one, or vice versa. The matrix $U_j$ is obtained from the eigenvectors associated with the most significant eigenvalues $\lambda_i$ of the covariance matrix $S_j$ of the data set $x_j$.

This training stage uses a number of measurements $K$ from the input data set representing different track conditions for the class $\alpha_j$. Every feature vector is then described by $\dot{x}_{kj}$ (4), $\dot{x}_{kj} \in \mathbb{R}^8$ (8 currents). Then the mean vector $\psi_j$ (5), the zero-mean one $x_{kj}$ (6) and the covariance matrix $S_j$ (7) are calculated.

$$\dot{x}_{kj} = \left[ I_{1e}^{(e} \ I_{1i}^{(e} \ I_{2i}^{(e} \ I_{2e}^{(e} \ I_{1e}^{(r} \ I_{1i}^{(r} \ I_{2i}^{(r} \ I_{2e}^{(r} \right]_{kj} \quad k = \{1, …, K\}; j = \{1, …, c\} \tag{4}$$

$$\psi_j = \frac{1}{K} \sum_{k=1}^{K} \dot{x}_{kj} \tag{5}$$

$$x_{kj} = \dot{x}_{kj} - \psi_j \tag{6}$$

$$S_j = \frac{1}{K} \sum_{k=1}^{K} (x_{kj})(x_{kj})^T \tag{7}$$

Fig. 6 represents the described off-line phase for a class $\alpha_j$. The same process has to be carried out for every class.

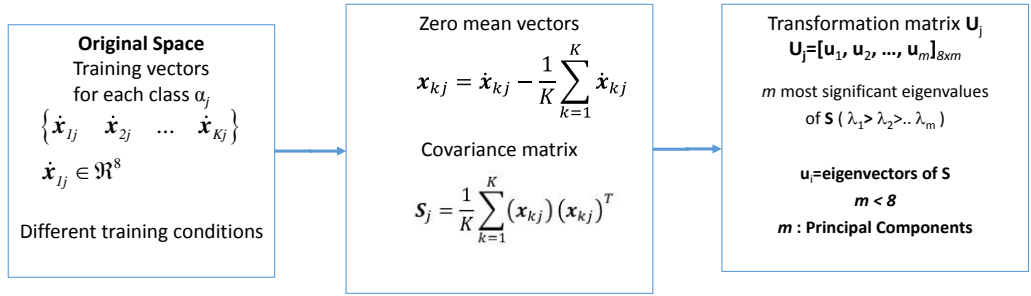

Fig. 6. PCA off-line process for class $\alpha_j$ (training phase).

The second stage is on-line, known as classification phase. By using the transformation matrix $U_j$, the feature vector $\dot{x}$ that is received from the measuring system is converted to a zero mean vector $x$ and is projected onto each transformed space (one per class $\alpha_j$) to obtain the new transformed feature vector $y_j$, according to (8).

$$y_j = U_j^T x \tag{8}$$

Later the reconstruction is computed by using (9), obtaining the reconstructed vector $\hat{x}_j$ per class $\alpha_j$.

$$\hat{x}_j = U_j \, y_j \tag{9}$$

The reconstructed information $\hat{x}_j$ differs from the original one $x$ in different magnitudes depending on the similarity level that exists between the new data $x$ and those which were used to obtain the transformation matrix $U_j$. This difference is known as the reconstruction error $\mathcal{E}_j$ (10), and it is computed for every class $\alpha_j$ by using the Mahalanobis distance between $x$ and $\hat{x}_j$.

$$\mathcal{E}_j = (x - \hat{x}_j)^T S_j^{-1} (x - \hat{x}_j) \tag{10}$$

Then, the minimum reconstruction error $\mathcal{E}_j$ classifies the input vector $\dot{x}$ as a membership of the class $\alpha_j$. Fig. 7 summarizes the described online process to classify a set of measurements.

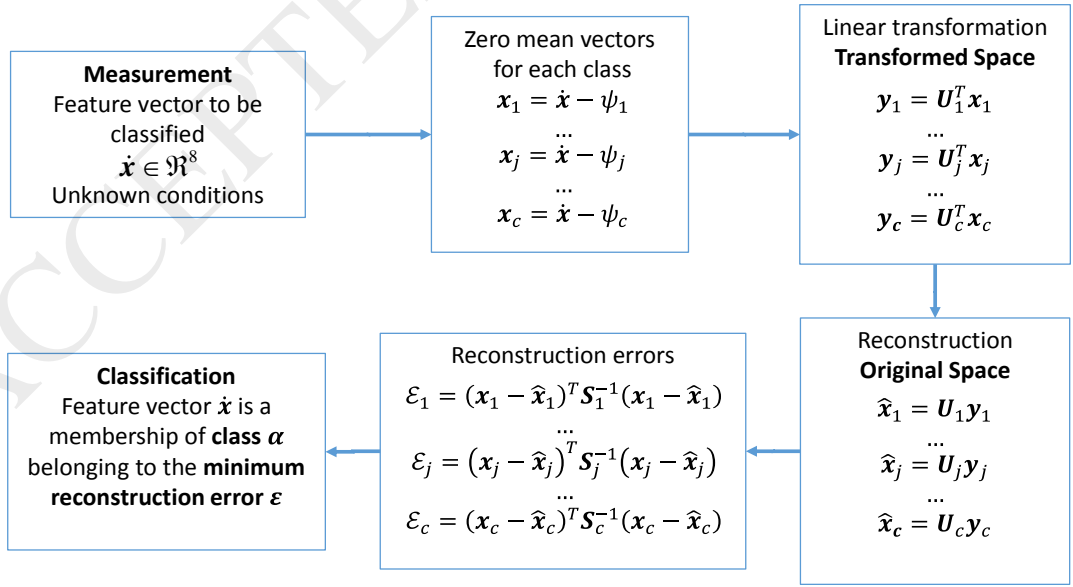

Fig. 7. PCA on-line process. Classification phase.

### 4.2. Railway breakage detection based on PCA

The global detection process has been already described in Section 3, divided into different phases with a set of classes for each phase (see Fig. 5) identifying the corresponding breakage. In order to be consistent with the description carried out in Section 4.1, each class is denoted as $\alpha_j$. For the independent injection there are only two classes per track, whereas, for the joint injection, the number of classes for the second and third phases is six and twelve, respectively.

  a) Phase 1

Table 4 shows the classes $\alpha_j$ considered in this phase, taking into account no correlation between tracks (independent injection). The analysis of classes $\{\alpha_1, \alpha_2\}$ is carried out independently of classes $\{\alpha_3, \alpha_4\}$.

| Class $\alpha_j$ | Class description | Breakage classification |
|---|---|---|
| $\alpha_1$ | $\alpha_{1,OK}$ | Track 1 is not broken |
| $\alpha_2$ | $\alpha_{1,BR}$ | Track 1 is broken |
| $\alpha_3$ | $\alpha_{2,OK}$ | Track 2 is not broken |
| $\alpha_4$ | $\alpha_{2,BR}$ | Track 2 is broken |

Table 4. Classes in the first detection phase (independent injection).

  b) Phase 2

This phase is reached when the result of the previous phase is a broken track (either class $\alpha_2$ or $\alpha_4$ or both of them). Now the objective is to find out which the broken rail is by using joint injection. Table 5 shows the possible classes $\{\alpha_1, \alpha_2, \alpha_3, \alpha_4, \alpha_5, \alpha_6\}$.

| Class $\alpha_j$ | Class description $\alpha_{t,r}$ | Breakage classification |
|---|---|---|
| $\alpha_1$ | $\alpha_{1,i}$ | Track 1, internal rail |
| $\alpha_2$ | $\alpha_{1,e}$ | Track 1, external rail |
| $\alpha_3$ | $\alpha_{1,ie}$ | Track 1, both rails |
| $\alpha_4$ | $\alpha_{2,i}$ | Track 2, internal rail |
| $\alpha_5$ | $\alpha_{2,e}$ | Track 2, external rail |
| $\alpha_6$ | $\alpha_{2,ie}$ | Track 2, both rails |

Table 5. Classes in the second detection phase (joint injection).

  c) Phase 3

If there exists only one breakage in the cell under study, the detection algorithm proceeds with the third phase. Once the track and the broken rail are identified, this phase tries to locate the breakage along the cell. For that purpose, and according to the developed hardware simulator (see subsection 2.2), three different locations have been considered: close to the emitting node, close to the receiving one, or in the intermediate area. Twelve breakage classes can be then identified, and the relationship between a class $\alpha_j$ and the associated breakage (see Fig. 3) is pointed out in Table 6. Note that, without loss of generality, more breakage locations might be considered, but the higher the number of classes the more complex the hardware simulator and, what is worst, the more difficult to get evidences in a real scenario.

| Class $\alpha_j$ | Class description $\alpha_{t,r,z}$ | Breakage label (Fig. 3) | Class description | | |
|---|---|---|---|---|---|
| | | | Track | Rail | Position of the breakage |
| $\alpha_1$ | $\alpha_{1,e,ne}$ | R1e 1/4 | 1 | External | Close to the emitting node, 1/4 [2km] |
| $\alpha_2$ | $\alpha_{1,i,ne}$ | R1i 1/4 | 1 | Internal | Close to the emitting node, 1/4 [2km] |
| $\alpha_3$ | $\alpha_{2,e,ne}$ | R2i 1/4 | 2 | Internal | Close to the emitting node, 1/4 [2km] |
| $\alpha_4$ | $\alpha_{2,i,ne}$ | R2e 1/4 | 2 | External | Close to the emitting node, 1/4 [2km] |
| $\alpha_5$ | $\alpha_{1,e,in}$ | R1e 2/4 | 1 | External | In the intermediate area, 2/4 [4km] |
| $\alpha_6$ | $\alpha_{1,i,in}$ | R1i 2/4 | 1 | Internal | In the intermediate area, 2/4 [4km] |
| $\alpha_7$ | $\alpha_{2,e,in}$ | R2i 2/4 | 2 | Internal | In the intermediate area, 2/4 [4km] |
| $\alpha_8$ | $\alpha_{2,i,in}$ | R2e 2/4 | 2 | External | In the intermediate area, 2/4 [4km] |
| $\alpha_9$ | $\alpha_{1,e,nr}$ | R1e 3/4 | 1 | External | Close to the receiving node, 3/4 [6km] |
| $\alpha_{10}$ | $\alpha_{1,i,nr}$ | R1i 3/4 | 1 | Internal | Close to the receiving node, 3/4 [6km] |
| $\alpha_{11}$ | $\alpha_{2,e,nr}$ | R2i 3/4 | 2 | Internal | Close to the receiving node, 3/4 [6km] |
| $\alpha_{12}$ | $\alpha_{2,i,nr}$ | R2e 3/4 | 2 | External | Close to the receiving node, 3/4 [6km] |

Table 6. Class description with regard to the third phase of the classification process (joint injection).

### 4.3. Separation of classes

In order to check the feasibility of the proposed classification process, it is required to analyze if the classes are separable. For that, a previous study has been carried out working on the hardware simulator. Through the simulator, one hundred transmissions have been performed for each one of the twelve classes $\alpha_j$ of the last phase, providing the corresponding measurements in the sensors. The main variations are due to the signal source noise, measurement noise and correlation process. Fig. 8 shows the 2D representation of the twelve classes $\alpha_j$ according to the auto-correlation value from two currents in the feature vector, namely: $I_{2i}^{(r}$ and $I_{1i}^{(r}$. Table 7 shows the statistical parameters related to the autocorrelation values of such currents (standard deviation, $\sigma_x$; mean, $\mu_x$; and the index of dispersion, $D_x = \sigma_x^2/\mu_x$), being $x$ and $y$ the autocorrelations of $I_{2i}^{(r}$ and $I_{1i}^{(r}$ respectively. The index of dispersion shows that the set of data per class is under-dispersed, confirming that the considered classes are separable.

| Class | $\sigma_x$ | $\mu_x$ | $D_x$ | $\sigma_y$ | $\mu_y$ | $D_y$ |
|---|---|---|---|---|---|---|
| R1e 1/4 | 50,92 | 19040 | 0,13 | 40,67 | 14065 | 0,12 |
| R1i 1/4 | 52,99 | 20191 | 0,14 | 35,56 | 11719 | 0,11 |
| R2i 1/4 | 64,46 | 21640 | 0,20 | 37,23 | 12874 | 0,11 |
| R2e 1/4 | 49,43 | 16327 | 0,15 | 28,01 | 9797 | 0,08 |
| R1e 2/4 | 54,83 | 19115 | 0,15 | 55,31 | 18131 | 0,17 |
| R1i 2/4 | 57,72 | 21076 | 0,16 | 29,05 | 11253 | 0,07 |
| R2i 2/4 | 74,14 | 24620 | 0,22 | 39,40 | 12930 | 0,12 |
| R2e 2/4 | 38,83 | 13902 | 0,11 | 28,67 | 8891 | 0,09 |
| R1e 3/4 | 57,45 | 20162 | 0,16 | 85,07 | 24668 | 0,30 |
| R1i 3/4 | 61,44 | 21409 | 0,17 | 18,19 | 6561 | 0,05 |
| R2i 3/4 | 71,81 | 29312 | 0,17 | 34,15 | 11606 | 0,10 |
| R2e 3/4 | 28,28 | 7505 | 0,11 | 28,27 | 8672 | 0,09 |

Table 7. Statistical parameters of the autocorrelation values of $I_{2i}^{(r}$ and $I_{1i}^{(r}$.

**4.4. Principal components**

According to the process depicted in Fig. 6, the feature vector $\dot{x}_{kj}$ is composed of the measured currents. Then, the eigenvectors of the matrix $U_j$ have been computed for each class of the three detection stages. The selected *m* eigenvectors correspond to the *m* most significant eigenvalues $\lambda_i$ of the covariance matrix $S_j$. The value of *m* is determined according to the criterion of getting a normalized residual root mean square error (RMSE) lower than 10% [30], defining RMSE as (11):

$$RMSE(m) = \frac{\sum_{i=m+1}^{n} \lambda_i}{\sum_{i=1}^{n} \lambda_i} \quad (11)$$

where *n* is the number of the original eigenvectors of the covariance matrix $S_j$.

The third classification stage is the most complex situation as the number of breakage classes is twelve. The PCA classifier has been trained by software, using the electrical railway model of the hardware simulator (dry track [42]) and different SNRs. To evaluate the effect of SNR, the double-track section has been simulated adding Gaussian white noise to the measurement of currents. The following SNR values have been considered: -10dB, -3dB, 0dB, +3dB and +10dB), this way 100 measurements have been obtained, thus providing *K*=500 training vectors $\dot{x}_{kj}$ per class $\alpha_j$. For all the analyzed situations the criterion of RMSE is fulfilled for *m*=4.

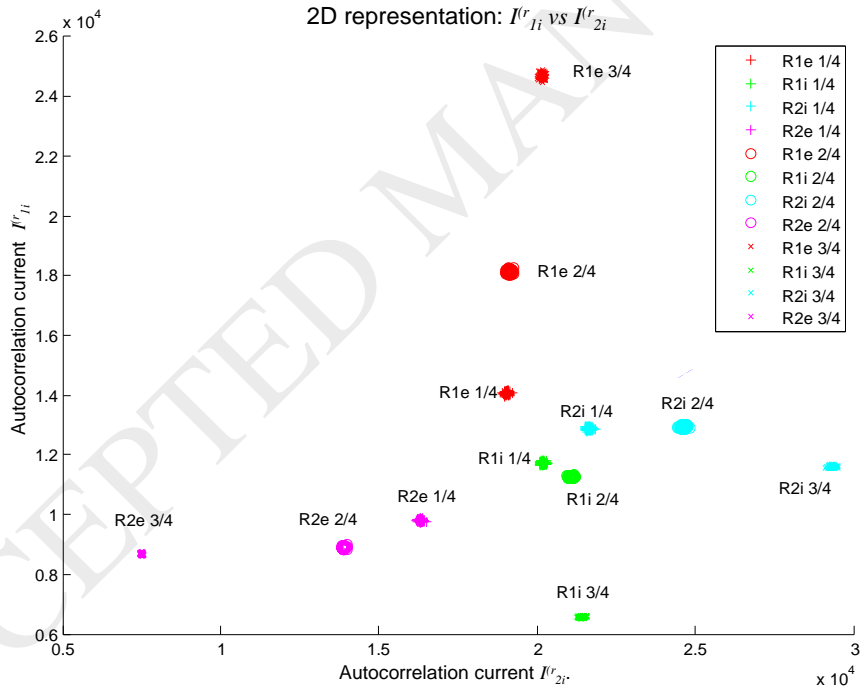

Fig. 8. 2D representation of the twelve classes $\alpha_j$ related to the third phase, according to the feature vector components $I_{1i}^{(r}$ and $I_{2i}^{(r}$.

As an example, Fig. 9 depicts the evolution of RMSE depending on *m* for the situation of "no broken rails" in track no. 1 (see Table 4: $\alpha_1 \equiv \alpha_{1,OK}$). The *m* principal eigenvectors correspond to the measurements provided by the sensors in the receiving node, $\left[ I_{1e}^{(r} \; I_{1i}^{(r} \; I_{2i}^{(r} \; I_{2e}^{(r} \right]$, as Table 8 shows.

| Current/sensor | Eigenvalue |
|---|---|
| $I_{1e}^{(e}$ | 10962,00 |
| $I_{1i}^{(e}$ | 13377,12 |
| $I_{2i}^{(e}$ | 15264,30 |
| $I_{2e}^{(e}$ | 28151,90 |
| $I_{1e}^{(r}$ | **96724,82** |
| $I_{1i}^{(r}$ | **125065,94** |
| $I_{2i}^{(r}$ | **433733,11** |
| $I_{2e}^{(r}$ | **593724,04** |

Table 8. Eigenvalues $\lambda_i$ of the covariance matrix $\mathbf{S_j}$.

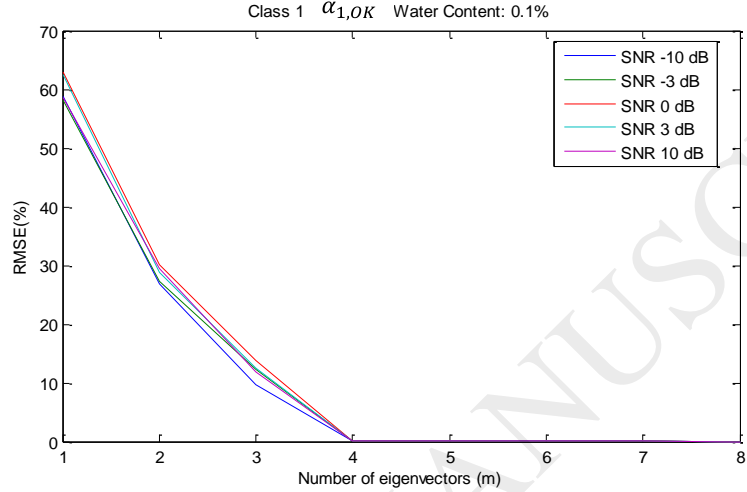

Fig. 9. Evolution of RMSE for class $\alpha_{1,OK}$ (no broken rail) for different SNRs and dry track.

## 5. Experimental results

As has been previously stated, it is very difficult to check the algorithm in a real environment. Nevertheless, thanks to the 8km long track hardware simulator different breakages can be generated. We would like to recall that when the state of the tracks is similar to one of the trained classes it is expected a low reconstruction error (as stated in Section 4.1). Depending on the detection phase, the number and type of classes is different: from Track OK (phase 1) to a breakage in a particular position of one rail (phase 3) (see Figure 5). If the detection algorithm proceeds with the third phase it means that there is only one breakage. Then the classifier provides the most likely position of such breakage according to the data used in the training phase. This section shows a variety of cases that validate the proposed algorithm, by using the principal components indicated in Section 4.4 (see Table 8).

### 5.1. Examples of classification with one breakage

In this case a breakage of the external rail of track no. 1 close to the receiving node (R1e3/4 according to Fig. 3) has been generated. The results obtained at every detection phase are the following:

a) First phase. It is carried out with individual injection for each track *t* (see Fig. 2a). Table 9 shows the reconstruction errors $\mathcal{E}_j$ provided by the PCA classifier.

| Class $\alpha_j$ | Class description $\alpha_{t,r}$ | Error $\mathcal{E}_j$ |
|---|---|---|
| $\alpha_1$ | $\alpha_{1,OK}$ | 9920,45 |
| $\alpha_2$ | $\alpha_{1,BR}$ | **35,82** |
| $\alpha_3$ | $\alpha_{2,OK}$ | **1,44** |
| $\alpha_4$ | $\alpha_{2,BR}$ | 31673,52 |

Table 9. Reconstruction errors $\mathcal{E}_j$ for classes in the first detection phase.

According to the obtained reconstruction errors, only track no. 1 is broken ($\alpha_{1,BR}$ $\alpha_{2,OK}$).

b) Second phase. If there is only one breakage, using joint injection (see Fig. 2b), it can be concluded on which rail is the breakage. Table 10 shows the reconstruction errors $\mathcal{E}_j$, concluding that the most likely situation is that the external rail is broken.

c) Third phase. Now the objective is to locate the breakage in one of the three trained positions. According to the results shown in Table 11, the most likely class is $\alpha_9$, which means that the broken rail is the external one in track no. 1, close to the receiving node (R1e3/4, see Fig. 3).

Tables 12 and 13 show other situations of broken rail when the third phase is reached. In Table 12 the broken rail is the external one in track no. 1, close to the emitting node; in Table 13, the broken rail is the internal one in track no. 2, close to the receiving node.

| Class $\alpha_j$ | Class description $\alpha_{t,r}$ | Error $\mathcal{E}_j$ |
|---|---|---|
| $\alpha_1$ | $\alpha_{1,i}$ | 10336,22 |
| $\alpha_2$ | $\alpha_{1,e}$ | **38,62** |
| $\alpha_3$ | $\alpha_{1,ie}$ | 6839,42 |
| $\alpha_4$ | $\alpha_{2,i}$ | 5615,59 |
| $\alpha_5$ | $\alpha_{2,e}$ | 62785,48 |
| $\alpha_6$ | $\alpha_{2,ie}$ | 59650,45 |

Table 10. Reconstruction errors $\mathcal{E}_j$ for classes in the second detection phase.

| Class $\alpha_j$ | Identification Label | Error $\mathcal{E}_j$ |
|---|---|---|
| $\alpha_1$ | R1e 1/4 | 15420,17 |
| $\alpha_2$ | R1i 1/4 | 44587,84 |
| $\alpha_3$ | R2i 1/4 | 30467,62 |
| $\alpha_4$ | R2e 1/4 | 54025,33 |
| $\alpha_5$ | R1e 2/4 | 9266,05 |
| $\alpha_6$ | R1i 2/4 | 35769,79 |
| $\alpha_7$ | R2i 2/4 | 23893,04 |
| $\alpha_8$ | R2e 2/4 | 77888,31 |
| $\alpha_9$ | R1e 3/4 | **0,16** |
| $\alpha_{10}$ | R1i 3/4 | 20532,88 |
| $\alpha_{11}$ | R2i 3/4 | 40110,36 |
| $\alpha_{12}$ | R2e 3/4 | 12486,60 |

Table 11. Classification results when there is a breakage in track no. 1, external rail, close to the receiving node.

| Class $\alpha_j$ | Identification Label | Error $\mathcal{E}_j$ |
|---|---|---|
| $\alpha_1$ | R1e 1/4 | **2,17** |
| $\alpha_2$ | R1i 1/4 | 5273,75 |
| $\alpha_3$ | R2i 1/4 | 2885,13 |
| $\alpha_4$ | R2e 1/4 | 5292,65 |
| $\alpha_5$ | R1e 2/4 | 2657,95 |
| $\alpha_6$ | R1i 2/4 | 4908,01 |
| $\alpha_7$ | R2i 2/4 | 1715,28 |
| $\alpha_8$ | R2e 2/4 | 15632,48 |
| $\alpha_9$ | R1e 3/4 | 34183,80 |
| $\alpha_{10}$ | R1i 3/4 | 42068,59 |
| $\alpha_{11}$ | R2i 3/4 | 8189,46 |
| $\alpha_{12}$ | R2e 3/4 | 59228,44 |

Table 12. Classification results when there is a breakage in track no. 1, external rail, close to the emitting node.

| Class $\alpha_j$ | Identification Label | Error $\mathcal{E}_j$ |
|---|---|---|
| $\alpha_1$ | R1e 1/4 | 22084,76 |
| $\alpha_2$ | R1i 1/4 | 7734,30 |
| $\alpha_3$ | R2i 1/4 | 5904,91 |
| $\alpha_4$ | R2e 1/4 | 20240,75 |
| $\alpha_5$ | R1e 2/4 | 15555,27 |
| $\alpha_6$ | R1i 2/4 | 10731,98 |
| $\alpha_7$ | R2i 2/4 | 4145,56 |
| $\alpha_8$ | R2e 2/4 | 41044,40 |
| $\alpha_9$ | R1e 3/4 | 94499,30 |
| $\alpha_{10}$ | R1i 3/4 | 23973,33 |
| $\alpha_{11}$ | R2i 3/4 | **0,40** |
| $\alpha_{12}$ | R2e 3/4 | 17285,23 |
| $\alpha_{13}$ | R1e 1/4 | 22084,76 |

Table 13. Classification results when there is a breakage in track no. 2, internal rail, close to the receiving node.

### 5.2. Examples of classification with more than one breakage

In this second scenario, a case with several broken rails is analyzed. As was explained in Section 3, if the state of tracks is OK, or a breakage is detected in both tracks, the classification process finishes at the first phase. If the breakage belongs to one track but in both rails, the process finishes at the second phase. So, hereinafter, a case with a double breakage in track no. 1 is described, placed at both rails in the intermediate area (R1i2/4 and R1e2/4).

a) First phase. Table 14 shows the reconstruction error $\mathcal{E}_j$ for every track.

| Class $\alpha_j$ | Class description $\alpha_{t,r}$ | Error $\mathcal{E}_j$ |
|---|---|---|
| $\alpha_1$ | $\alpha_{1,OK}$ | 9529,88 |
| $\alpha_2$ | $\alpha_{1,BR}$ | **97,016** |
| $\alpha_3$ | $\alpha_{2,OK}$ | **426,14** |
| $\alpha_4$ | $\alpha_{2,BR}$ | 13569,02 |

Table 14. Classification results with a double breakage in track no. 1.

b) Second phase. Table 15 shows the classification results for track no. 1, discarding the results from track no. 2 since this is not broken. The most likely class is the one for a simultaneous breakage in both rails. The classification process finishes in this phase because there is more than one breakage, and the system is not able to locate them.

| Class $\alpha_j$ | Class description $\alpha_{t,r}$ | Error $\mathcal{E}_j$ |
|---|---|---|
| $\alpha_1$ | $\alpha_{1,i}$ | 27686,90 |
| $\alpha_2$ | $\alpha_{1,e}$ | 66724,82 |
| $\alpha_3$ | $\alpha_{1,ie}$ | **512,15** |
| $\alpha_4$ | $\alpha_{2,i}$ | 33218,59 |
| $\alpha_5$ | $\alpha_{2,e}$ | 17219,67 |
| $\alpha_6$ | $\alpha_{2,ie}$ | 43219,21 |

Table 15. Detection of a double breakage in track no. 1.

## 6. Discussion

Due to the difficulty of collecting a significant number of rail breakages for each class in a real double track railway, the authors' proposal has been validated on a hardware simulator, specifically designed for a dry railway and an 800 Hz working frequency.

Although our railway prototype does not include all the conditions that can affect the track impedance (e.g. geological properties, environmental conditions), its behavior is very similar to modern high-speed railway lines, currently built on a ballast platform and sleepers in such a way that the rail-ground conductivity and permeability are minimized. In our opinion, the main constraint of this study is to cope with high-moisture conditions, due to it reduces the electrical insulation among rails.

To estimate the effect of water content in soil on the characteristic impedance of the tracks [41] [42], and therefore on the capability to detect breakages, a software has been developed. Furthermore, this software facilitates the simulation of different levels of noise on the signals measured in the emitting and receiving nodes, thus evaluating the effect of different signal-to-noise ratios (see Section 4.4). Two situations of water content have been checked: dry double-track (water content 0.1%) and wet double-track (water content=1%) for the 8km long section. The characteristic impedance of the tracks, R-L-C-G values, has been modelled according to [42], considering in this way two different models in the classification process. For real situations, it should be obtained a batch of models according to different water content ranges and, once the current ballast water content is measured, the suitable model should be used for the classification process. A similar solution is proposed in [43].

Moreover, the SNR effect in Fig. 9, Fig. 10 confirms that the criterion of RMSE for the mentioned water content is fulfilled with $m$=4 (RMSE<10%).

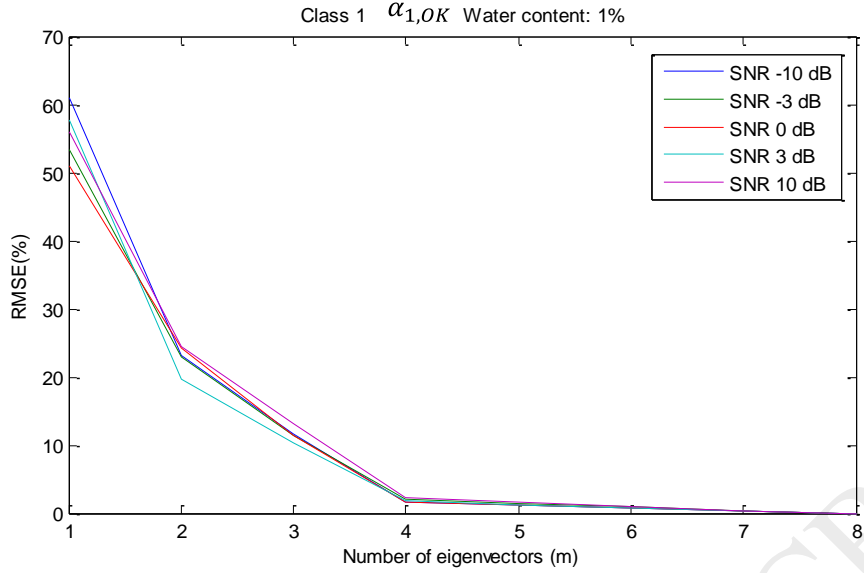

Fig. 10. Evolution of RMSE for class $\alpha_{1,OK}$ (no broken rail) for different SNRs and wet track.

As the results show, the different breakages are correctly classified assuming that the most likely class is the one with the minimum reconstruction error $\mathcal{E}_j$. We have empirically checked with the hardware simulator that the selected class has always a reconstruction error that is lower than 5% of any of the remaining reconstruction errors, providing a high robustness to the classification process. However, there can exist real situations with an unusual low reconstruction error that can provoke a wrong classification. In such a case, we should consider the $T^2$ statistic to confirm the result provided by the classification stage [37]. If the value of the $T^2$ statistic is lower than a threshold stablished for a confidence level [37][43], we can conclude that the sample under analysis has a low variation within the model. As an example, Figure 11 shows the $T^2$ statistic for three classes: a) there is no breakage; b) breakage at track 1, external rail close to the emitting node, 1/4 [2km] (Class R1e1/4); c) breakage at track 2, internal rail close to the receiving node, 3/4 [6km] (Class R2i3/4). For this example, we have used 100 test samples per each situation that were not used in the training stage. The dashed line in each plot shows a confidence interval of 95%, showing that there are a few outliers.

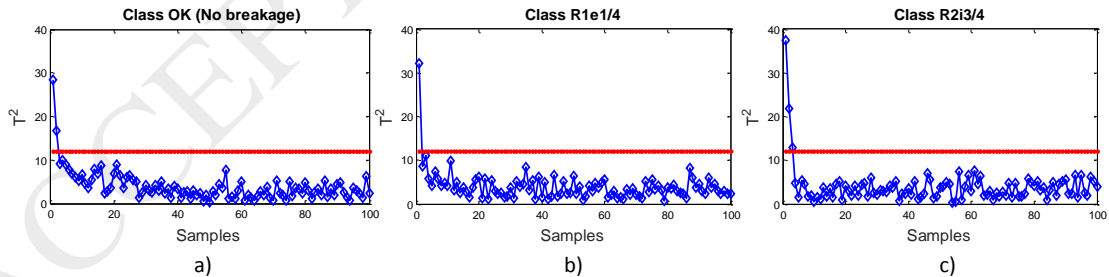

Fig. 11. $T^2$ statistic (threshold for a 95% confidence interval): a) Tracks are OK; b) Breakage at track 1, external rail close to the emitting node, 1/4 [2km] (Class R1e1/4); c) Breakage at track 2, internal rail close to the receiving node, 3/4 [6km] (Class R2i3/4).

As can be derived from the analysis of Figure 11, the values of $T^2$ show a low dispersion. The reason of these results is explained in [37][43]. The $T^2$ statistic is associated to systematic errors of the process, and the data used in this work have a negligible systematic component. This is mainly due to the electronic system [38] that has been designed for avoiding this problem, and we are using a double track hardware simulator, free of systematic errors as well. Note that the electronic equipment includes a self-diagnosis process which runs before each new breakage detection. In this way, any electronic subsystem, including the sensory one, is checked and any systematic variation is detected in advance. The self-diagnosis also includes a system calibration stage in order to determine the suitable level balance of the eight currents in no breakage conditions.

As was previously indicated, it is required to obtain a set of models for each class at each real work condition in order to get a robust classifier. As the real environment is unknown, there can be extreme situations for which the established classes may not be separable. Then, the $T^2$ statistic should be calculated in any situation to guarantee that the data fit the used model. Nevertheless, the electronic system designed for every node [38] is able to detect them and report about such circumstances, thus generating the corresponding alarm (the classifier is not in operation).

In the current work we have not considered the case of more than one breakage per rail although our hardware simulator is able to generate such situations. For a future work, a new improvement would consist in including more than one breakage per rail and per zone in the simulator and in analyzing how the designed classifier works.

## 7. Conclusions

A PCA-based strategy of rail breakage detection for a double-track railway line has been presented here for monitoring systems and maintenance of railway lines, which avoids the use of vehicles to carry out this task, and requires a reduced number of devices to be located in the infrastructure. The detection is performed through several phases. The proposal gradually discriminates the situations of broken rail, identifying the track and the damaged rail(s). If there is only one breakage, the system is able to report the most likely area where that breakage could be along the 8km long track section, thus distinguishing three possible locations: close to the emitting node, close to the receiving one, and in the intermediate area.

The proposed classification algorithm uses four principal components from the feature vector corresponding to the measured currents at each rail in the emitting and receiving nodes. The algorithm has been previously trained by using a double-track railway line software simulator, including different SNR conditions and water content in soil. The experimental validation has been conducted on a double-track hardware simulator that allows twelve different breakages. Experimental results show a 100% success rate identifying the breakages. Furthermore, if the hardware simulator included more breakage switches, the number of breakage locations could be increased.

The main constraint of the proposal for being implemented in a real railway infrastructure is to have enough breakage records for each one of the analyzed classes. Nevertheless, from the point of view of the monitoring and the maintenance of the railway line, the result provided by the first detection phase (whether there is a broken rail or not) is relevant enough to safety control the railway traffic. Note that for that purpose it is only necessary to get available records of the tracks without breakages.

**Acknowledgements**

This work has been supported by the Instalaciones Inabensa Company (SD3 Research Project), and we would like to thank the kindly help of J.C. Cortes, R. Arévalo and L. Turmo.

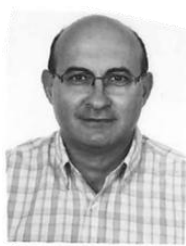

**Felipe Espinosa** (M'98-SM'15) received the M. Eng. degree from the Polytechnic University of Madrid, Madrid, Spain, in 1991 and the Ph.D. degree from the University of Alcala (UAH), Alcalá de Henares, Spain, in 1999, both in Telecommunication. In 2000, he became an Associate Professor with the GEINTRA Group, Department of Electronics, UAH. His current research interests include network control systems, wireless sensor networks and event-based control applied to intelligent transportation systems and industrial automation.

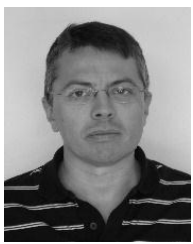

**J. Jesús García** (M'06) obtained his B.Sc. degree in Electronic Engineering from the University of Alcala (Spain) in 1992, and his M.Sc. degree from Polytechnic University of Valencia (Spain) in 1999. He received the Ph.D. degree with distinction from University of Alcala (Spain) in 2006. He is currently an Associate Professor of Digital and Analog Electronic at the Electronics Department of the University of Alcala. His research areas are multi-sensor integration, local positioning systems and sensory systems for railway safety.

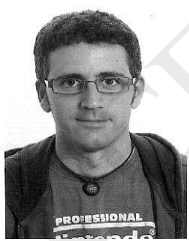

**Álvaro Hernández** (M'06-SM'15) obtained his PhD from the University of Alcala (Spain), and from the Blaise Pascal University (France) in 2003. He is currently an associate professor of Digital Systems and Electronic Design at the Electronics Department in the University of Alcala. His research areas are multi-sensor integration, electronic systems for mobile robots, digital and embedded systems.

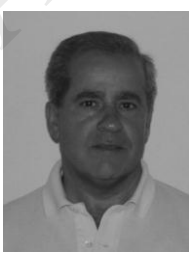

**Manuel Mazo** received a PhD degree in Telecommunications in 1988, and the M.Sc in Eng. Degree in Telecommunications in 1982, both from the Polytechnical University of Madrid (Spain). He is currently a full Professor in the Electronics Department, University of Alcala, Alcalá de Henares, Madrid, Spain. His research interests include electronics control, intelligent sensors, robot sensing and perception, intelligent spaces, electronics systems for railway safety, and wheelchairs for physically disabled people.

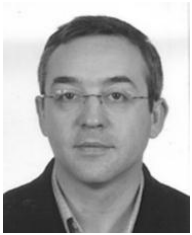

**Jesús Ureña** received the B.S. degree in Electronics Engineering and the M.S. degree in Telecommunications Engineering from the Polytechnical University of Madrid, Madrid, Spain, in 1986 and 1992, respectively; and the Ph.D. degree in Telecommunications from the University of Alcala, Madrid, Spain, in 1998. Since 1986, he has been with the Department of Electronics, University of Alcala, currently as a Professor. His current research interests are in the areas of low-level ultrasonic signal processing, local positioning systems (LPSs), and sensory systems for railway infrastructure.

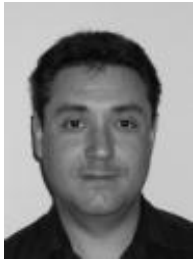

**José A. Jiménez** obtained his Electronic Telecommunication Engineering degree in 1996 from the University of Valencia (Spain). In 2004, he obtained his PhD from the University of Alcala (Spain). He is currently an Associate Professor of Ins trumentation Electronics in the Electronics Department at the University of Alcala. His research areas are multi-sensor integration, sensory systems applied to robotic, instrumentation and electronic systems for mobile robots.

**Ignacio Fernández** received the Ph.D. degree in Telecommunication from the University of Alcala (UAH), Spain, in 2005. He is currently an Associate Professor in the Electronics Department of the University of Alcala, involved in hardware and software electronics design. His research interests are focused on intelligent spaces and intelligent transportation systems.

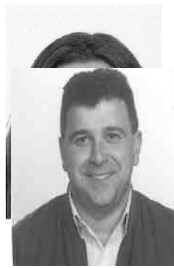

**Mª Carmen Pérez** (M'07) received the M. S. degree in Electronics Engineering and the PhD degree from the University of Alcala (UAH), Spain, in 2004 and 2009, respectively. She is currently an Assistant Professor at the Electronics Department from the University of Alcala. Since 2003 she has collaborated on several research projects in the areas of sequence design, low-level ultrasonic signal processing and computing architectures.

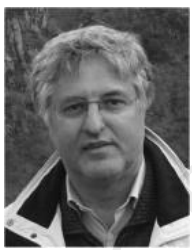

**Juan C. García** is PhD in Telecommunications and special award from the University of Alcala in 2001, having obtained the Telecommunication Eng. degree from the UPM (Madrid). After various activities in private industry, shared with the university, he is an Associate Professor at the University of Alcala (UAH) since 1990. His field of work focuses on field and service robotics, mainly in the area of Assistive Technology. His areas of interest include Computer Vision, Embedded Systems, Wireless Sensors and Systems, Systems Integration and Human Machine Interfaces. Currently he is a member of the Group of Intelligent Spaces and Transportation (GEINTRA) from UAH.